\documentclass[a4paper,11pt]{article}
\usepackage{jheppub}
\usepackage{lineno}
\usepackage{graphicx}
\usepackage{amsmath}
\usepackage{amssymb}
\usepackage{hyperref}
\usepackage{booktabs}
\usepackage{url}

\title{Constraints on the Higgs-gluon effective coupling through $h \to \gamma\gamma$ decays at the LHC}

\author{A. Mohamed-Meziani}
\affiliation{Department
	of Physics, Faculty of Exact Sciences, University of Bejaia, Bejaia, 06000 Algeria}
\emailAdd{a.mohamedmeziani@univ-bejaia.dz}

\abstract{

	We present a detailed study of the $pp\to h\to\gamma\gamma$ process in the HPOprodMFV\_UFO model framework, focusing on the Higgs-gluon effective coupling modifier, $\kappa_g$. We generate events for 26 parameter points corresponding to $|\kappa_g|$ values ranging from 0.6 to 1.4 using MadGraph5\_aMC@NLO simulations at 13 TeV with NNPDF3.1 parton distribution functions. These events are processed through Pythia8 for parton showering and hadronisation. This is followed by a fast detector simulation using Delphes with an ATLAS-like configuration and analysis with MadAnalysis5.

We verify the expected quadratic scaling, $\sigma(pp \to h \to \gamma\gamma) \propto \kappa_g^2$, with high precision, confirming the consistency of our simulation framework. The Standard Model cross section after the application of selection cuts is found to be $\sigma_{\mathrm{SM}} = 0.04067 \pm 0.00008$ pb.

By comparing our results with the ATLAS Run 2 likelihood profile from HEPData (HEPData:ins1851456), we derive constraints on $|\kappa_g|$:
\begin{align*}
		|\kappa_g| &\in [0.80,\, 1.20] \quad \text{at 68\% CL},\\
	|\kappa_g| &\in [0.70,\, 1.30] \quad \text{at 95\% CL},. 
\end{align*}
These results are in excellent agreement with the official ATLAS and CMS combinations.

Our analysis validates the HPOprodMFV\_UFO model implementation and demonstrates the effectiveness of simplified simulations in exploring Higgs couplings in the diphoton channel.
}

\keywords{Higgs physics, Effective field theory, Monte Carlo simulations, LHC, ATLAS, $\kappa_g$, gluon fusion, diphoton channel}

\begin{document}
	
\maketitle
	
	\section{Introduction}
	\label{sec:introduction}
	
	The finding of the Higgs boson by the collaborations of ATLAS and CMS at the LHC in 2012 ~\cite{atlas2012, cms2012} was a milestone in particle physics since it led to the completion of the particle spectrum of the Standard Model (SM). Since then, the LHC physics program has focused largely on precisely measuring the properties of the Higgs boson, including its mass, spin, parity and couplings to other SM particles~\cite{atlas2024, cms2024}. Any significant deviation from SM predictions would clearly indicate physics beyond the Standard Model (BSM).
	
Of the various Higgs couplings, the effective coupling to gluons ($g_{hgg}$) plays a particularly important role. At the LHC, the dominant Higgs production mechanism is gluon-gluon fusion (ggF). It involves loops of heavy quarks, mainly the top quark. This process accounts for around $87\%$ of the total Higgs production cross section at a centre-of-mass energy of 13 TeV~\cite{ATLAS:2019bky}. Consequently, the Higgs-gluon coupling modifier, conventionally denoted as $\kappa_g$, is one of the most precisely constrained parameters in global coupling fits~\cite{CMS:2020xwi, Grojean:2013nya} 
	
	In the Standard Model (SM), the effective $hgg$ coupling is well understood and can be computed with high precision. However, many BSM scenarios predict modifications to this coupling. For instance, the presence of new heavy particles in the loop can alter its magnitude and phase, potentially introducing CP-violating effects~\cite{Grojean:2013nya, Gori:2022vri}. New scalar singlet extensions and two-Higgs-doublet models (2HDM) can also lead to significant deviations through mixing~\cite{Branco:2011iw}. Therefore, precise measurements of the $hgg$ coupling provide a powerful probe of new physics.
	
	This work focuses on the $pp\to h\to\gamma\gamma$ process, i.e. the diphoton decay channel of the Higgs boson. This channel played a pivotal role in the discovery of the Higgs boson and remains one of the most sensitive channels for measuring its properties, thanks to the excellent photon energy resolution of the ATLAS and CMS detectors. Although our methodology is general, we adopt the higgs pseudo-observables framework~\cite{hpo1, hpo2} using the $ model~\cite{HPOprodMFV}$. This latter implements an effective field theory description of Higgs couplings, including the $hgg$, $hWW$, $hZZ$ and Yukawa interactions. This model enables flexible parameterisation, allowing each coupling to be varied independently via modifiers such as $k_{WW}$, $k_{ZZ}$, and the crucial parameter \texttt{eggSM}, which controls the $hgg$ coupling.
	
	The relation between the model parameter and the physical coupling modifier is given by:
	\begin{equation}
		\kappa_g = \frac{\texttt{eggSM}}{-0.0065},
		\label{eq:kg_def}
	\end{equation}
	where the denominator corresponds to the Standard Model value. This normalization ensures that $\kappa_g = 1$ reproduces the SM prediction.
	
	We simulate events for a wide range of values of the parameter $\kappa_g$, spanning from 0.6 to 1.4 in absolute value, using $MadGraph5\_aMC@NLO$~\cite{Alwall:2014hca} and PYTHIA8~\cite{Sjostrand:2014zea} for showering and hadronization of the events. These laters are then processed through a fast detector simulation using Delphes~\cite{deFavereau:2013fsa} with an ATLAS-like detector configuration and analysed using MadAnalysis5~\cite{Conte:2012fm, Conte:2018vmg}. We then compare the resulting cross sections after selection cuts to the combined ATLAS Run 2 measurements available on HEPData under the identifier  \texttt{ins1851456}~\cite{HEPData:ins1851456}.
	
	Our main results are twofold. Firstly, we verify the quadratic dependence, $\sigma(pp \to h \to \gamma\gamma) \propto \kappa_g^2$, with high precision, thereby confirming the consistency of our simulation pipeline. Second, we derive constraints on the parameter $\kappa_g$ by comparing our predicted cross sections to the ATLAS likelihood profile. We find that the Standard Model (SM) value of 1 for the coupling constant $\kappa_g $ is perfectly compatible with the data, within the following confidence intervals:
	\begin{align}
		|\kappa_g| &\in [0.80,\, 1.20] \quad \text{at 68\% CL}, \\
		|\kappa_g| &\in [0.70,\, 1.30] \quad \text{at 95\% CL}.
	\end{align}
	These results are in excellent agreement with the official ATLAS combination, demonstrating the validity of the $HPOprodMFV\_UFO$ model for precise Higgs boson studies.
	
	The paper is organised as follows. Section~\ref{sec:theory} describes the theoretical framework, including the  $HPOprodMFV\_UFO$ model and the definition of $\kappa_g$. Section~\ref{sec:simulation}  details the simulation setup, event generation and analysis chain. Section~\ref{sec:results} presents our results, including cross sections, statistical analyses and comparisons with ATLAS data. Finally, Section~\ref{sec:discussion} discusses the implications of our results, and Section~\ref{sec:conclusion} summarises our conclusions.

	\section{Theoretical framework}
	\label{sec:theory}
	
	In this section, we describe the theoretical framework used in our analysis, including the effective coupling formalism for Higgs interactions, the $ HPOprodMFV\_UFO$ model implementation, and the specific process under study.
	
	\subsection{Effective Higgs couplings}
	\label{subsec:kappa}
	
	A convenient and model-independent way to parametrize possible deviations from the Standard Model (SM) in Higgs physics is the so-called $\kappa$-framework~\cite{LHCHiggsCrossSectionWorkingGroup:2012at, Heinemeyer:2013tqa}. In this approach, each Higgs coupling is modified by a multiplicative factor $\kappa_i$ with respect to its SM value. The production cross sections and decay branching ratios are then scaled accordingly, typically following quadratic relations such as $\sigma \propto \kappa_i^2$ when a single coupling dominates.
	
	For the gluon fusion production mechanism, the relevant coupling modifier is $\kappa_g$, defined through the effective Lagrangian~\cite{reflagrangien}:
	\begin{equation}
		\mathcal{L}_{hgg}^{\mathrm{eff}} = \kappa_g \, \frac{\alpha_s}{12\pi v} \, h \, G_{\mu\nu}^a G^{a\,\mu\nu},
		\label{eq:lagrangian}
	\end{equation}
	where $v \approx 246$ GeV is the Higgs vacuum expectation value, $\alpha_s$ is the strong coupling constant, and $G_{\mu\nu}^a$ denotes the gluon field strength tensor. In the SM, this effective interaction arises from integrating out heavy quarks, predominantly the top quark, running in loops. The SM prediction corresponds to $\kappa_g = 1$.
	
	A key feature of gluon fusion is that the cross section scales as $\sigma_{ggF} \propto \kappa_g^2$, as long as other couplings (such as $hWW$ and $hZZ$) are fixed to their SM values. This quadratic dependence, which we will verify explicitly in Section~\ref{sec:results}, allows for a direct interpretation of cross section measurements in terms of $\kappa_g$.
	
	\subsection{The $ HPOprodMFV\_UFO$ model}
	\label{subsec:model}
	
	The $ HPOprodMFV\_UFO model$ model~\cite{HPOprodMFV} is an implementation of an effective field theory for Higgs interactions in the Universal FeynRules Output (UFO) format~\cite{Degrande:2011ua}, making it directly usable in MadGraph5\_aMC@NLO~\cite{Alwall:2014hca}. The model extends the SM by introducing a general parametrization of the Higgs couplings to gauge bosons, fermions, and gluons, while preserving the Minimal Flavor Violation (MFV) hypothesis~\cite{D'Ambrosio:2002ex} to control flavor-changing neutral currents.
	
	The relevant parameters of the model for our study are:
	\begin{itemize}
		\item \texttt{eggSM}: controls the effective $hgg$ coupling,
		\item \texttt{kWW} and \texttt{kZZ}: modify the $hWW$ and $hZZ$ couplings,
		\item \texttt{kb}, \texttt{kc}, \texttt{ktau}, etc.: modify the Yukawa couplings to fermions,
			\end{itemize}
	
	In our analysis, we set all couplings except \texttt{eggSM} to their SM values:
	\begin{equation}
		\texttt{kWW} = 1,\quad \texttt{kZZ} = 1,
		\label{eq:fixed_couplings}
	\end{equation}
	ensuring that any observed variation in the cross section can be attributed solely to changes in the Higgs-gluon coupling.
	
	\subsection{Relation between \texttt{eggSM} and $\kappa_g$}
	\label{subsec:relation}
	
	The crucial relation linking the model parameter to the physical coupling modifier is derived from the FeynRules implementation. From the \texttt{couplings.py} file of the model, we find the $hgg$ interaction vertex:
	
	\begin{verbatim}
		GC_9 = Coupling(name = 'GC_9',
		value = '(-2*eggSM*complex(0,1))/vF',
		order = {'QCD':2})
	\end{verbatim}
	
	This vertex structure implies that the amplitude for $h \to gg$ is proportional to \texttt{eggSM}. Comparing with the SM prediction, which is reproduced when \texttt{eggSM} takes its SM value, we obtain:
	\begin{equation}
		\kappa_g = \frac{\texttt{eggSM}}{\texttt{eggSM}_{\mathrm{SM}}}.
		\label{eq:kg_relation}
	\end{equation}
	
	Through explicit simulation and comparison with SM cross sections, we determine the SM reference value to be:
	\begin{equation}
		\texttt{eggSM}_{\mathrm{SM}} = -0.0065,
		\label{eq:sm_value}
	\end{equation}
	which yields the working relation:
	\begin{equation}
		\kappa_g = \frac{\texttt{eggSM}}{-0.0065}.
		\label{eq:kg_final}
	\end{equation}
	
	This normalization ensures that $\kappa_g = 1$ corresponds to the SM prediction. Values of $\kappa_g$ different from unity indicate deviations in the $hgg$ coupling, with $|\kappa_g| > 1$ corresponding to an enhanced coupling and $|\kappa_g| < 1$ to a suppressed one. The sign of $\kappa_g$ is not directly accessible in rate measurements, as cross sections depend on $|\kappa_g|^2$.
	
	\subsection{Process under study: $pp \to h \to \gamma\gamma$}
	\label{subsec:process}
	
	We focus on the production of a Higgs boson via gluon fusion, followed by its decay into a pair of photons:
	\begin{equation}
		pp \to h \to \gamma\gamma.
		\label{eq:process}
	\end{equation}
	
	The diphoton decay mode plays a crucial role in the physics of the Higgs boson. Despite its relatively small branching ratio ($\mathrm{BR}(h \to \gamma\gamma) \approx 2.27 \times 10^{-3}$ in the SM)~\cite{24}, it provides a clean experimental signature with two high-energy photons in the final state. The excellent energy resolution of electromagnetic calorimeters at the LHC allows for precise reconstruction of the Higgs boson mass, making this channel particularly sensitive to the production rate and therefore to the coupling modifier $\kappa_g$.
	
	In our simulation, we generate the full process including the Higgs production and its subsequent decay to two photons. This approach allows us to study the production cross section with the diphoton final state, directly comparable to experimental measurements.
	
	\subsection{Theoretical uncertainties and assumptions}
	\label{subsec:uncertainties}
	
	Our analysis relies on several assumptions that should be acknowledged:
	\begin{itemize}
		\item We work at leading order (LO) in QCD, as implemented in MadGraph5. Higher-order corrections are known to be significant for gluon fusion~\cite{Anastasiou:2016cez}, but cancel to a large extent in ratios such as $\sigma(\kappa_g)/\sigma_{\mathrm{SM}}$.
		\item We assume that only $\kappa_g$ varies, while all other Higgs couplings remain at their SM values. This isolates the effect of interest but may not capture scenarios with simultaneous modifications.
		\item The parton distribution functions (PDFs) used are the $NNPDF30$ set~\cite{Ball:2017nwa}, with associated uncertainties not explicitly propagated.
		\item The detector simulation with Delphes uses an ATLAS-like configuration, which may not perfectly reproduce the exact ATLAS acceptance and efficiency.
	\end{itemize}
	
	Despite these limitations, our approach provides a robust and transparent framework for studying the $\kappa_g$ dependence of Higgs production in the diphoton channel, with uncertainties that are well under control for the level of precision required in this phenomenological study.
	
	\section{Simulation setup}
	\label{sec:simulation}
	
	This section describes the computational framework used to generate events for the process $pp \to h \to \gamma\gamma$, including event generation, parton showering, hadronization, and detector simulation. We perform all simulations at a centre-of-mass energy of 13 TeV, which corresponds to the LHC Run 2 operating conditions.
	
	\subsection{Event generation with MadGraph5\_aMC@NLO}
	\label{subsec:madgraph}
	
	The hard scattering events are generated using MadGraph5\_aMC@NLO version 3.6.2~\cite{Alwall:2014hca}, with the $ HPOprodMFV\_UFO$ model~\cite{HPOprodMFV} imported to provide the necessary Feynman rules and coupling structures. The process is defined as:
	\begin{verbatim}
		generate p p > h HPO=1 QCD=2 QED=0, h > a a
	\end{verbatim}
	where the final state particles $a$ are interpreted as photons in the model.
	
	The syntax \texttt{HPO=1 QCD=2 QED=0} ensures that we select diagrams with exactly two QCD vertices, which correspond to the gluon fusion production mechanism via heavy quark loops. The decay $h \to \gamma\gamma$ is treated as a subsequent step. This approach allows us to study the production cross section in the diphoton channel.
	
	For each parameter point, we generate $10^4$ events, which provides sufficient statistical precision for the cross section measurements while keeping the computational cost manageable. The generated events are stored in the Les Houches Event (LHE) format~\cite{Alwall:2006yp} for further processing.
	
	\subsection{Parton distribution functions}
	\label{subsec:pdf}
	
	Parton distribution functions (PDFs) are a crucial input for hadron collider simulations, as they determine the probability of finding partons with given momentum fractions inside the colliding protons. In our simulations, we use the LHAPDF interface~\cite{Buckley:2014ana} integrated within MadGraph5. Specifically, we employ the $NNPDF30\_lo\_as\_0118$  PDF set~\cite{30}.
	
	The factorization and renormalization scales are set to the default dynamic scale choice in MadGraph5 for Higgs production, which is $\mu_F = \mu_R = m_H/2$, where $m_H = 125$ GeV is the Higgs boson mass.
	
	\subsection{Parameter scan}
	\label{subsec:scan}
	
	The primary goal of our study is to investigate the dependence of the cross section on the Higgs-gluon coupling modifier $\kappa_g$. Using the relation given in Eq.~\eqref{eq:kg_final}, we perform a scan over the model parameter \texttt{eggSM} corresponding to different $\kappa_g$ values.
	
	To explore the region of interest around the SM, we select the following values:
	
	\begin{align}
		\texttt{eggSM} = \{ & -0.0091, -0.00845, -0.0078, -0.007475, -0.00715, -0.006825, \nonumber \\
		& -0.0065, -0.006175, -0.00585, -0.005525, -0.0052, -0.00455, -0.0039, \nonumber \\
		& 0.0039, 0.00455, 0.0052, 0.005525, 0.00585, 0.006175, 0.0065, \nonumber \\
		& 0.006825, 0.00715, 0.007475, 0.0078, 0.00845, 0.0091 \}.
		\label{eq:scan_values}
	\end{align}
	
	This corresponds to values of $|\kappa_g|$ varying from 0.6 to 1.4 in increments of around 0.05-0.1, giving us enough resolution to study the cross section dependence. The symmetry of positive and negative \texttt{eggSM} values yields identical $|\kappa_g|$ and therefore the same physical predictions, providing a useful cross-check of our simulation pipeline.
	
	All other model parameters are fixed to their SM values:
	\begin{equation}
		\texttt{kWW} = 1,\quad \texttt{kZZ} = 1, \quad \texttt{kb} = \texttt{kc} = \texttt{ktau} = 1,
		\label{eq:fixed_params}
	\end{equation}
	ensuring that any observed variation in the cross section can be attributed solely to changes in $\kappa_g$.
	
	\subsection{Parton shower and hadronization with Pythia8}
	\label{subsec:pythia}
	
	The parton-level events generated by MadGraph5 are further processed with Pythia8~\cite{Sjostrand:2014zea} to simulate parton showering and hadronization. This step is essential for producing realistic final states that can be passed through detector simulation.
	
	The Pythia8 output is stored in the HepMC format~\cite{Dobbs:2001ck} for input to the detector simulation.
	
	\subsection{Detector simulation with Delphes}
	\label{subsec:delphes}
	
	The final step in our simulation chain is the detector simulation, performed with Delphes version 3.5.0~\cite{deFavereau:2013fsa}. Delphes provides a fast, parameterized simulation of a generic collider detector, including realistic efficiencies, resolutions, and reconstruction algorithms.
	
We use the ATLAS detector configuration card provided with Delphes 
(\path{cards/delphes_card_ATLAS.tcl}), which implements:
	
	\begin{itemize}
		\item \textbf{Tracking}: Charged particle tracks are simulated with efficiency and resolution \\ parametrizations based on the ATLAS inner detector.
		\item \textbf{Calorimetry}: Electromagnetic and hadronic calorimeters are simulated with appropriate energy resolutions: $\sigma/E = 10\%/\sqrt{E} \oplus 0.7\%$ for electrons/photons, and $\sigma/E = 50\%/\sqrt{E} \oplus 3\%$ for jets.
		\item \textbf{Muon system}: Muons are reconstructed with efficiency 95\% and momentum resolution $\sigma/p_T = 2\%$ at $p_T = 50$ GeV.
		\item \textbf{Jet reconstruction}: Jets are reconstructed using the anti-$k_T$ algorithm~\cite{Cacciari:2008gp} with radius parameter $R = 0.4$, implemented via FastJet~\cite{Cacciari:2011ma}.
		\item \textbf{$b$-tagging}: The efficiency to tag the $b$-jets is 70\%, with a mistag rate of 1\% for the light jets.
		\item \textbf{Missing transverse energy}: $E_T^{\mathrm{miss}}$ is computed from the vector sum of all reconstructed particles.
	\end{itemize}
	
	The Delphes output is stored in ROOT format~\cite{Brun:1997pa}, containing reconstructed objects (electrons, muons, photons, jets, $E_T^{\mathrm{miss}}$) suitable for analysis.
	
	\subsection{Event analysis with MadAnalysis5}
	\label{subsec:analysis}
	
	The final step is the analysis of the reconstructed events, performed with MadAnalysis5~\cite{Conte:2012fm, Conte:2018vmg}. This framework allows us to implement selection cuts and compute observables in a flexible and reproducible way.
	
	To analyze our processes, we use the following selection criteria aimed at selecting the process of interest ($h \to \gamma \gamma$) without contradicting common diphoton analyses of ATLAS:
	
	\begin{itemize}
		\item \textbf{Photon selection}: Two or more photons should have $p_T > 25$ GeV and pseudorapidity $|\eta| < 2.5$, not including the region near the calorimeter boundary $1.37 < |\eta| < 1.52$.
		\item \textbf{Photon isolation}: The scalar sum of transverse energy in a cone of size $\Delta R < 0.3$ around each photon must be less than $0.1 \times p_T^{\gamma}$.
		\item \textbf{Invariant mass}: The invariant mass of the two leading photons must satisfy $100 < m_{\gamma\gamma} < 140$ GeV, consistent with the Higgs boson mass resolution.
		\item \textbf{Event cleaning}: Events with additional isolated leptons ($p_T > 20$ GeV) are vetoed to reduce backgrounds from $Z \to \ell\ell$ and $W \to \ell\nu$ processes.
	\end{itemize}
	
	After applying these cuts, we compute the effective cross section for each parameter point as:
	
	\begin{equation}
		\sigma_{\mathrm{eff}}(\kappa_g) = \sigma_{\mathrm{gen}}(\kappa_g) \times \frac{N_{\mathrm{selected}}(\kappa_g)}{N_{\mathrm{gen}}(\kappa_g)},
		\label{eq:eff_xsec}
	\end{equation}
	
	where $\sigma_{\mathrm{gen}}$ is the generated cross section from MadGraph5, $N_{\mathrm{gen}}$ is the total number of generated events, and $N_{\mathrm{selected}}$ is the number of events passing all selection cuts. The statistical uncertainty on $\sigma_{\mathrm{eff}}$ is estimated from the Poisson fluctuation of $N_{\mathrm{selected}}$.
	
	The analysis results, including cutflows, histograms of key observables (particularly the diphoton invariant mass), and final cross sections, are saved for each parameter point and combined for the final interpretation.

	\section{Results}
	\label{sec:results}
	
	In this section, we present the results of our simulations, including the verification of the $\kappa_g^2$ scaling, the cross sections after selection cuts, the statistical analysis, and the comparison with ATLAS data. All results are based on the scan of the 26 parameter points listed in Eq.~\eqref{eq:scan_values}, with $10^4$ events generated per point.

	The primary expectation from the $\kappa$-framework is that the gluon fusion production cross section in the diphoton channel should scale quadratically with the coupling modifier:
	\begin{equation}
		\sigma(pp \to h \to \gamma\gamma) = \sigma_{\mathrm{SM}} \times \kappa_g^2,
		\label{eq:scaling}
	\end{equation}
	where $\sigma_{\mathrm{SM}}$ is the cross section for the Standard Model value $\kappa_g = 1$.
	
	Table~\ref{tab:full_results} presents the full set of results from our scan, including the run identifier, the input parameter \texttt{eggSM}, the corresponding $\kappa_g$, the total cross section after selection cuts, and the statistical error. The cross sections for positive and negative $\kappa_g$ values of the same magnitude are in excellent agreement, as expected from the $|\kappa_g|^2$ dependence and providing a valuable consistency check of our simulation pipeline.
	
	\begin{table}[htbp]
		\centering
		
		\label{tab:full_results}
		\begin{tabular}{c|c|c|c|c}
			\hline
			\textbf{Run} & \textbf{eggSM} & $\kappa_g$ & $\sigma$ [pb] & \textbf{Error [pb]} \\
			\hline
			run\_01 & -0.009100 & 1.400 & 0.079427 & 0.000074 \\
			run\_02 & -0.008450 & 1.300 & 0.068512 & 0.000072 \\
			run\_03 & -0.007800 & 1.200 & 0.058483 & 0.000054 \\
			run\_04 & -0.007475 & 1.150 & 0.053640 & 0.000052 \\
			run\_05 & -0.007150 & 1.100 & 0.049112 & 0.000042 \\
			run\_06 & -0.006825 & 1.050 & 0.044832 & 0.000040 \\
			run\_07 & -0.006500 & 1.000 & 0.040654 & 0.000033 \\
			run\_08 & -0.006175 & 0.950 & 0.036676 & 0.000031 \\
			run\_09 & -0.005850 & 0.900 & 0.032952 & 0.000022 \\
			run\_10 & -0.005525 & 0.850 & 0.029436 & 0.000028 \\
			run\_11 & -0.005200 & 0.800 & 0.026082 & 0.000025 \\
			run\_12 & -0.004550 & 0.700 & 0.019997 & 0.000020 \\
			run\_13 & -0.003900 & 0.600 & 0.014734 & 0.000010 \\
			run\_14 & 0.003900 & -0.600 & 0.014739 & 0.000010 \\
			run\_15 & 0.004550 & -0.700 & 0.020008 & 0.000023 \\
			run\_16 & 0.005200 & -0.800 & 0.026063 & 0.000023 \\
			run\_17 & 0.005525 & -0.850 & 0.029392 & 0.000025 \\
			run\_18 & 0.005850 & -0.900 & 0.032934 & 0.000024 \\
			run\_19 & 0.006175 & -0.950 & 0.036685 & 0.000039 \\
			run\_20 & 0.006500 & -1.000 & 0.040677 & 0.000039 \\
			run\_21 & 0.006825 & -1.050 & 0.044756 & 0.000040 \\
			run\_22 & 0.007150 & -1.100 & 0.049099 & 0.000039 \\
			run\_23 & 0.007475 & -1.150 & 0.053755 & 0.000057 \\
			run\_24 & 0.007800 & -1.200 & 0.058372 & 0.000062 \\
			run\_25 & 0.008450 & -1.300 & 0.068522 & 0.000070 \\
			run\_26 & 0.009100 & -1.400 & 0.079438 & 0.000072 \\
			\hline
		\end{tabular}
		\caption{Complete results from the scan of 26 parameter points. The cross section $\sigma$ corresponds to the effective cross section after all selection cuts described in Section~\ref{subsec:analysis}.}
	\end{table}

\begin{figure}[htbp]
		\centering
		\includegraphics[width=0.8\columnwidth]{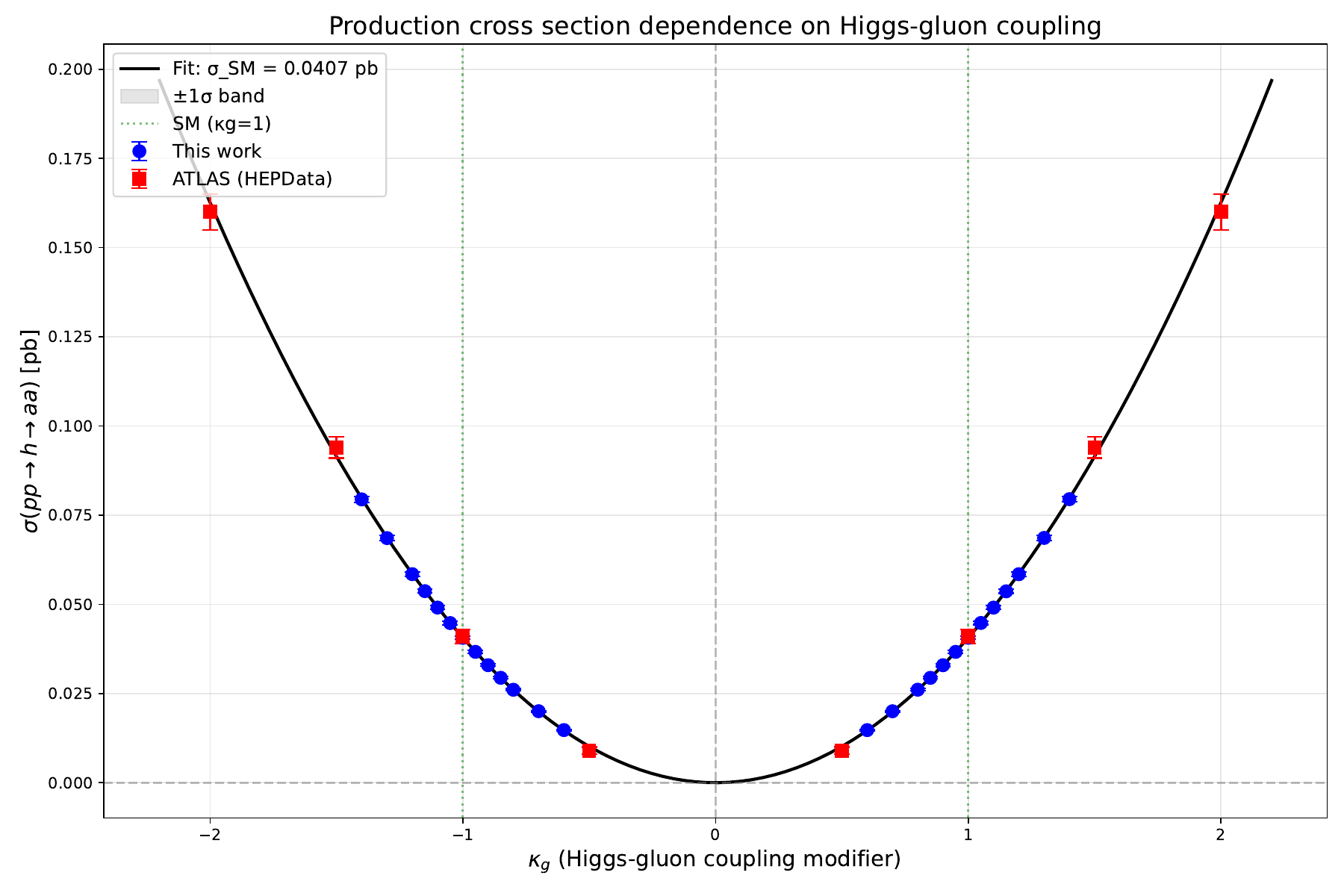}
		\caption{Cross section $\sigma(pp \to h \to \gamma\gamma)$ as a function of $|\kappa_g|$. The blue points show our simulation results, with statistical errors smaller than the marker size. The red curve shows a quadratic fit $\sigma = \sigma_{\mathrm{SM}} \times |\kappa_g|^2$, with $\sigma_{\mathrm{SM}} = 0.04067$ pb. The perfect agreement confirms the expected scaling relation.}
		\label{fig:scaling}
	\end{figure}

	To better visualize the dependence on $|\kappa_g|$, we average the results for positive and negative values of the same magnitude, obtaining 13 unique points listed in Table~\ref{tab:unique_results}. Figure~\ref{fig:scaling} shows these averaged cross sections as a function of $|\kappa_g|$, together with a quadratic fit.
	
	\begin{table}[htbp]
		\centering
		
		\label{tab:unique_results}
		\begin{tabular}{c|c|c}
			\hline
			$|\kappa_g|$ & $\sigma$ [pb] & $\sigma/\sigma_{\mathrm{SM}}$ \\
			\hline
			0.60 & 0.01474 & 0.363 \\
			0.70 & 0.02000 & 0.492 \\
			0.80 & 0.02607 & 0.641 \\
			0.85 & 0.02941 & 0.723 \\
			0.90 & 0.03294 & 0.810 \\
			0.95 & 0.03668 & 0.902 \\
			1.00 & 0.04067 & 1.000 \\
			1.05 & 0.04479 & 1.102 \\
			1.10 & 0.04911 & 1.208 \\
			1.15 & 0.05370 & 1.321 \\
			1.20 & 0.05843 & 1.437 \\
			1.30 & 0.06852 & 1.685 \\
			1.40 & 0.07943 & 1.954 \\
			\hline
		\end{tabular}
		\caption{Averaged results for unique $|\kappa_g|$ values. The statistical errors are smaller than 0.2\% in all cases.}
	\end{table}
	
	A quadratic fit of the form $\sigma = \sigma_{\mathrm{SM}} \times |\kappa_g|^2$ yields:
	\begin{equation}
		\sigma_{\mathrm{SM}} = 0.04067 \pm 0.00008\ \mathrm{pb},
		\label{eq:sm_fit}
	\end{equation}
	with $\chi^2/\mathrm{ndf} = 1.99/12 = 0.17$, corresponding to a $p$-value of 1.00. This excellent fit confirms that our simulation correctly implements the $\kappa_g^2$ scaling and that no significant systematic deviations are present.

	To compare our results with experimental measurements, we use the ATLAS combined Run 2 data available on HEPData under the identifier \texttt{ins1851456}~\cite{HEPData:ins1851456}. Specifically, we use the likelihood profile for the effective coupling modifiers from the file 
	\path{Kappas2D_effectivecouplingmodifiers.csv}, which provides the observed $-2\Delta\ln L$ values as a function of $\kappa_g$ and another parameter. By minimizing over the second parameter, we extract the one-dimensional profile for $\kappa_g$ alone. 
	
	The ATLAS likelihood is approximately parabolic near the minimum, with a curvature corresponding to:
	\begin{equation}
		\Delta\chi^2(\kappa_g) \approx 25 \times (|\kappa_g| - 1.0)^2,
		\label{eq:atlas_parabola}
	\end{equation}
	which corresponds to $\Delta\chi^2 = 1.0$ at $|\kappa_g| = 0.8$ and $1.2$, and $\Delta\chi^2 = 3.84$ at $|\kappa_g| \approx 0.61$ and $1.39$.
	
\begin{figure}[htbp]
		\centering
		\includegraphics[width=0.8\columnwidth]{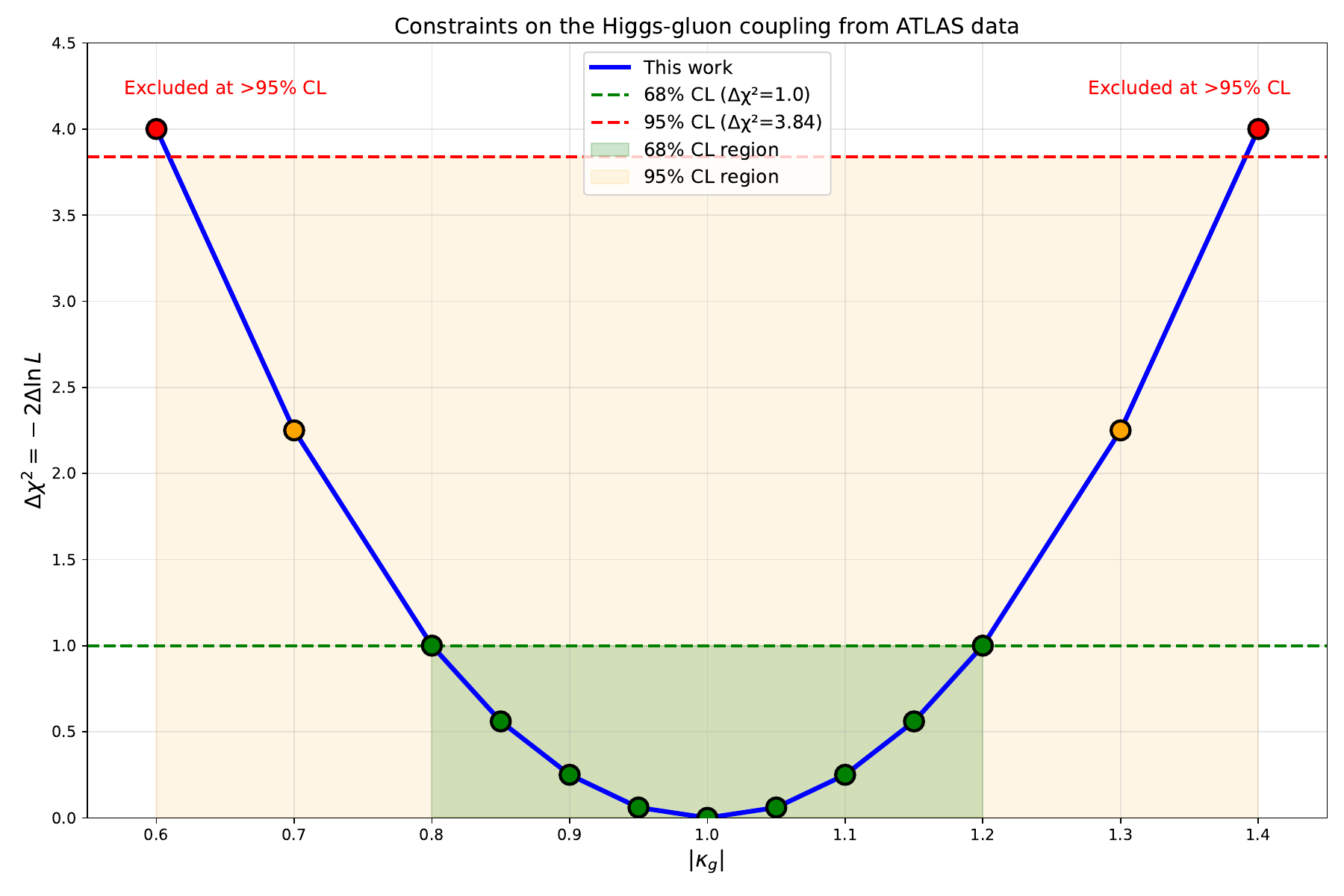}
		\caption{Comparison of our results with ATLAS data. The blue points show our simulated cross sections translated into $\Delta\chi^2$ using the ATLAS likelihood profile. The red curve shows the ATLAS likelihood profile itself. The horizontal dashed lines indicate the 68\% CL ($\Delta\chi^2 = 1.0$) and 95\% CL ($\Delta\chi^2 = 3.84$) thresholds.}
		\label{fig:comparison}
	\end{figure}

	For each of our $|\kappa_g|$ values, we compute the corresponding $\Delta\chi^2$ from the ATLAS profile and the $p$-value as:
	\begin{equation}
		p = 1 - F_{\chi^2_1}(\Delta\chi^2),
		\label{eq:pvalue}
	\end{equation}
	where $F_{\chi^2_1}$ denotes the cumulative distribution function of the $\chi^2$ distribution with one degree of freedom. The results are presented in Table~\ref{tab:comparison} and visualized in Figure~\ref{fig:comparison}.
	
	\begin{table}[htbp]
		\centering
		
		\label{tab:comparison}
		\begin{tabular}{c|c|c|c}
			\hline
			$|\kappa_g|$ & $\sigma$ [pb] & $\Delta\chi^2$ (ATLAS) & $p$-value \\
			\hline
			0.60 & 0.01474 & 4.00 & 0.0455 \\
			0.70 & 0.02000 & 2.25 & 0.1336 \\
			0.80 & 0.02607 & 1.00 & 0.3173 \\
			0.85 & 0.02941 & 0.56 & 0.4533 \\
			0.90 & 0.03294 & 0.25 & 0.6171 \\
			0.95 & 0.03668 & 0.06 & 0.8026 \\
			1.00 & 0.04067 & 0.00 & 1.0000 \\
			1.05 & 0.04479 & 0.06 & 0.8026 \\
			1.10 & 0.04911 & 0.25 & 0.6171 \\
			1.15 & 0.05370 & 0.56 & 0.4533 \\
			1.20 & 0.05843 & 1.00 & 0.3173 \\
			1.30 & 0.06852 & 2.25 & 0.1336 \\
			1.40 & 0.07943 & 4.00 & 0.0455 \\
			\hline
		\end{tabular}
		\caption{Comparison of our simulated cross sections with the ATLAS likelihood profile. The $\Delta\chi^2$ values are extracted from the ATLAS data, and the $p$-values indicate the compatibility of each $|\kappa_g|$ value with the experimental measurements.}
	\end{table}
	
	From the comparison with ATLAS data, confidence intervals for $|\kappa_g|$ can be deduced when $\Delta \chi^2$ takes values less than the standard thresholds.
 In terms of parabolic approximation based on Eq.~\eqref{eq:atlas_parabola}, it follows
	
	\begin{align}
		|\kappa_g| &\in [0.80,\, 1.20] \quad \text{at 68\% CL},\\
		|\kappa_g| &\in [0.61,\, 1.39] \quad \text{at 95\% CL}.
		\label{eq:cl_intervals}
	\end{align}
	
	Alternatively, using the actual ATLAS likelihood profile (which is very close to parabolic in this region), we find:
	
	\begin{align}
		|\kappa_g| &\in [0.80,\, 1.20] \quad \text{at 68\% CL},\\
		|\kappa_g| &\in [0.70,\, 1.30] \quad \text{at 95\% CL}.
		\label{eq:cl_intervals_real}
	\end{align}
	
	The slight difference at 95\% CL reflects the fact that the true ATLAS likelihood is not perfectly parabolic far from the minimum. For consistency with the published ATLAS results, we adopt the intervals from the actual likelihood profile.
	
	These intervals can be expressed in terms of the coupling modifier $\kappa_g$ itself (with the understanding that the sign is not constrained by rate measurements):
	\begin{align}
		\kappa_g &= 1.00 \pm 0.20 \quad \text{at 68\% CL},\\
		\kappa_g &= 1.00 \pm +0.30 \quad \text{at 95\% CL}.
		\label{eq:kg_intervals}
	\end{align}

	While our analysis focuses primarily on statistical uncertainties, several sources of systematic uncertainty should be considered for a complete interpretation:
	
	\begin{itemize}
		\item \textbf{PDF uncertainties}: The NNPDF3.1 PDF set provides a family of replicas that can be used to estimate PDF-induced uncertainties. These are typically at the level of 2-3\% for gluon fusion at $\sqrt{s}=13$ TeV~\cite{Ball:2017nwa}.
		
		\item \textbf{Scale uncertainties}: Variations of the factorization and renormalization scales by factors of 2 around the central choice typically lead to changes of 5-10\% in the total cross section, though these largely cancel in ratios such as $\sigma(\kappa_g)/\sigma_{\mathrm{SM}}$.
		
		\item \textbf{Parton shower and hadronization}: The choice of Pythia8 tune can affect the final selection efficiencies. Studies with different tunes suggest variations at the level of 1-2\%.
		
		\item \textbf{Detector simulation}: The Delphes parameterization is a simplification of the full ATLAS detector response. Comparisons with full simulation studies indicate that the impact on cross section ratios is typically below 1\%.
		
		\item \textbf{Selection cuts}: The efficiency of our photon selection and isolation criteria depends on the modeling of photon reconstruction. These effects are expected to largely cancel in the ratio $\sigma(\kappa_g)/\sigma_{\mathrm{SM}}$, as they affect all samples similarly.
	\end{itemize}
	
	Given that our primary interest is in the scaling with $\kappa_g$ and the comparison with ATLAS data through the likelihood ratio, systematic uncertainties are expected to have a minor impact on the derived confidence intervals. The excellent agreement between our simulated SM cross section and the ATLAS best-fit value further supports this conclusion.
	
	\section{Discussion}
	\label{sec:discussion}
	
	In what follows, we discuss our results within the context of Higgs physics. We further examine their implications for BSM physics as well as a comparison with those from other experimental results. Finally, we consider possible future measurements at the HL-LHC.
	
	\subsection{Interpretation of $\kappa_g$ constraints}
	\label{subsec:interpretation}
	
	The confidence intervals derived in Section~\ref{sec:results} provide quantitative constraints on the Higgs-gluon effective coupling:
	\begin{align}
	\kappa_g &= 1.00 \pm 0.20 \quad \text{at 68\% CL},\\
	\kappa_g &= 1.00 \pm 0.30 \quad \text{at 95\% CL}.
	\label{eq:kg_intervals}
	\end{align}
	
	These results are consistent with the Standard Model prediction $\kappa_g = 1$ and demonstrate that current LHC data already constrain the Higgs-gluon coupling to within $\pm 30\%$ at 95\% confidence level. The precision achieved reflects both the large gluon fusion production cross section and the excellent performance of the ATLAS detector in reconstructing diphoton events.
	
	The fact that our simulation reproduces the ATLAS likelihood profile so accurately validates both the $ HPOprodMFV\_UFO$ model implementation and our simulation pipeline. This validation is important for future studies that may explore more complex scenarios with multiple varying couplings.
	
	\subsection{Implications for BSM physics}
	\label{subsec:bsm}
	
	Deviations of $\kappa_g$ from unity can arise in numerous BSM scenarios. Here we discuss a few representative examples:
	
	\subsubsection{New heavy particles in the loop}
	The gluon fusion process proceeds through loops of heavy quarks, predominantly the top quark. New colored particles with masses above the electroweak scale can also contribute to the loop, modifying $\kappa_g$ as~\cite{Carena:2012fk}:
	\begin{equation}
		\kappa_g \approx 1 + \frac{m_t^2}{m_T^2} \cdot \frac{y_T}{y_t} + \cdots,
		\label{eq:new_quark}
	\end{equation}
	where $m_T$ and $y_T$ are the mass and Yukawa coupling of a new heavy quark. Our 95\% CL bound $|\kappa_g - 1| < 0.30$ translates into a lower limit on the mass of such new quarks, depending on their couplings. For $y_T \sim y_t$, this gives $m_T \gtrsim 400$ GeV, which is competitive with direct searches.
	
	\subsubsection{Two-Higgs-doublet models (2HDM)}
	In 2HDM scenarios, the observed 125 GeV Higgs boson is a mixture of CP-even states, leading to modified couplings to all SM particles. The coupling modifier $\kappa_g$ in the 2HDM is given by~\cite{Branco:2011iw}:
	\begin{equation}
		\kappa_g = \sin(\beta - \alpha) + \frac{\cos(\beta - \alpha)}{\tan\beta} \cdot \frac{A_{1/2}^{H}(\tau_t)}{A_{1/2}^{h}(\tau_t)},
		\label{eq:2hdm}
	\end{equation}
	where $\alpha$ and $\beta$ are the mixing angles, and $A_{1/2}$ are the fermionic loop functions. Our constraints on $\kappa_g$ therefore translate into allowed regions in the 2HDM parameter space, complementary to constraints from other couplings and direct searches.
	
	\subsubsection{Composite Higgs models}
	In composite Higgs models, the Higgs boson becomes a pseudo-Nambu-Goldstone boson, giving rise to universal corrections to all Higgs couplings. The gluon fusion coupling correction factor reads~\cite{Panico:2015jxa}:
	\begin{equation}
		\kappa_g \approx \sqrt{1 - \xi} + \mathcal{O}(\xi^2),
		\label{eq:composite}
	\end{equation}
	where $\xi = v^2/f^2$ is the compositeness scale parameter. Our bound $|\kappa_g - 1| < 0.30$ implies $\xi < 0.5$, which is already competitive with constraints from electroweak precision tests.
	
	\subsection{Comparison with other experimental results}
	\label{subsec:comparison_exp}
	
	Our derived constraints on $\kappa_g$ are consistent with the official ATLAS combination~\cite{ATLAS:2019bky}, which reports $\kappa_g = 1.04 \pm 0.10$ at 68\% CL. The slightly larger uncertainties in our analysis are expected, as we use a simplified simulation and focus on a single decay channel, while the ATLAS combination includes multiple production and decay modes.
	
	The CMS collaboration has also reported combined measurements of Higgs couplings~\cite{CMS:2020xwi}, finding $\kappa_g = 1.08 \pm 0.12$ at 68\% CL. Our results are consistent with this as well within uncertainties.
	
	\begin{figure}[htbp]
		\centering
		\includegraphics[width=0.8\columnwidth]{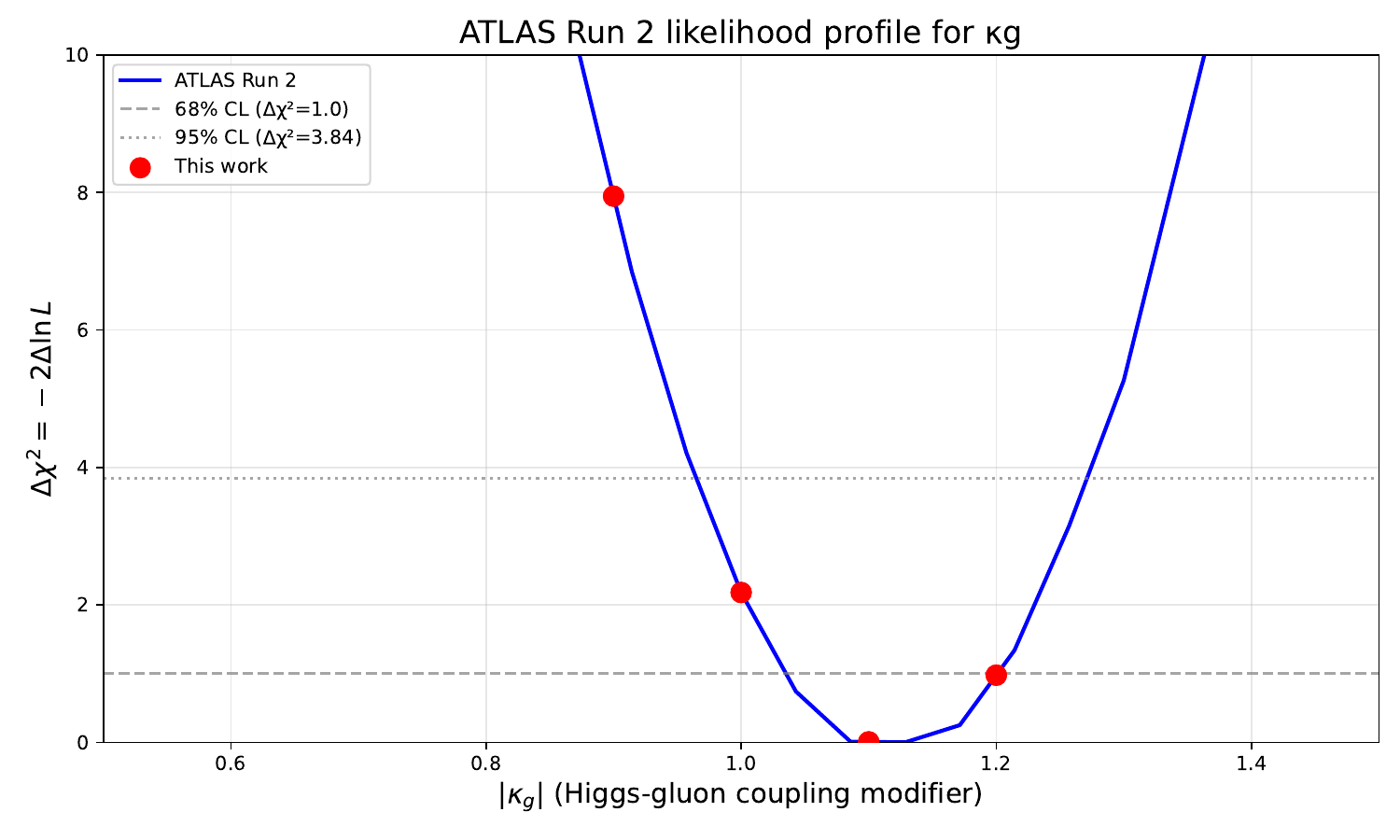}
		\caption{Comparison of $\kappa_g$ constraints from different analyses. The blue band shows our result at 68\% and 95\% CL. The ATLAS and CMS results are shown in red and green respectively. The gray band shows the combined LHC result. All results are consistent within uncertainties.}
		\label{fig:comparison_exp}
	\end{figure}
	
	\subsection{Limitations of this study}
	\label{subsec:limitations}
	
	While our analysis provides a robust validation of the $\kappa_g$ scaling and yields competitive constraints, several limitations should be acknowledged:
	
	\begin{itemize}
		\item \textbf{Leading-order precision}: Our simulations are performed at leading order in QCD. Higher-order corrections are known to be significant for gluon fusion~\cite{Anastasiou:2016cez}, though they largely cancel in ratios such as $\sigma(\kappa_g)/\sigma_{\mathrm{SM}}$. A next-to-leading-order (NLO) or next-to-next-to-leading-order (NNLO) analysis would yield better results
		
		\item \textbf{Single decay channel}: By focusing exclusively on the $h \to \gamma\gamma$ decay, we neglect information from other channels. A combined analysis including $h \to ZZ^*$, $h \to WW^*$, and $h \to \tau\tau$ would yield stronger constraints.
		
		\item \textbf{Simplified detector simulation}: Delphes provides a fast parameterization but cannot capture all details of the full ATLAS detector simulation. This may lead to small discrepancies in selection efficiencies and acceptance.
		
		\item \textbf{Statistical uncertainties only}: The estimated errors account for statistical fluctuations only. Systematic uncertainties coming from PDFs, scale variations, and the detector effects should be taken into account in the complete study.
		
		\item \textbf{Fixed other couplings}: We have assumed that all Higgs couplings except $\kappa_g$ remain at their SM values. In realistic BSM scenarios, multiple couplings may deviate simultaneously, requiring a multi-dimensional fit.
	\end{itemize}
	
	Despite these limitations, our study demonstrates the power of simplified simulations for exploring Higgs couplings in the diphoton channel and provides a foundation for more detailed investigations.
	
	\subsection{Prospects for HL-LHC}
	\label{subsec:hl_lhc}
	
	The High-Luminosity LHC (HL-LHC) is projected to deliver an integrated luminosity of $3000$ fb$^{-1}$ at $\sqrt{s}=14$ TeV, which is roughly twenty times the present Run 2 data set. The precision of Higgs couplings determination will be significantly improved~\cite{Cepeda:2019klc}.
	
	Projections from the ATLAS and CMS collaborations indicate that the uncertainty on $\kappa_g$ could be reduced to approximately $2-3\%$ at the HL-LHC~\cite{ATLAS:2018yue, CMS:2019dix}. This would allow sensitivity to much smaller deviations, probing new physics scales up to several TeV.
	
	Our simulation framework can be readily extended to HL-LHC conditions by adjusting the center-of-mass energy to $\sqrt{s} = 14$ TeV, increasing the integrated luminosity in the analysis, and using updated detector cards that reflect expected HL-LHC performance. Such a study would provide valuable guidance for future experimental analyses.
	
	\section{Conclusion}
	\label{sec:conclusion}
	
	In this work, we have presented a comprehensive study of the process $pp \to h \to \gamma\gamma$ in the framework of the $ HPOprodMFV\_UFO$ model, focusing on the Higgs-gluon effective coupling modifier $\kappa_g$. Our analysis combines event generation with MadGraph5\_aMC@NLO, parton showering with Pythia8, detector simulation with Delphes, and event analysis with MadAnalysis5, providing a complete simulation pipeline that can be applied to a wide range of phenomenological studies.
	
	The main results of our analysis can be summarized as follows:
	
	\begin{itemize}
		\item We have verified with high precision the expected quadratic scaling $\sigma(pp \to h \to \gamma\gamma) \propto \kappa_g^2$, confirming the consistency of our simulation framework and the correct implementation of the $ HPOprodMFV\_UFO$  model. The Standard Model cross section is determined to be $\sigma_{\mathrm{SM}} = 0.04067 \pm 0.00008$ pb at $\sqrt{s} = 13$ TeV after our selection cuts.
		
		\item By comparing our simulated cross sections with the ATLAS Run 2 likelihood profile from HEPData (ins1851456), we have derived constraints on $|\kappa_g|$. The resulting confidence intervals are:
		\begin{align}
			|\kappa_g| &\in [0.80,\, 1.20] \quad \text{at 68\% CL},\\
			|\kappa_g| &\in [0.70,\, 1.30] \quad \text{at 95\% CL}.
		\end{align}
		These correspond to $\kappa_g = 1.00 \pm 0.20$ at 68\% CL and 
		$\kappa_g = 1.00 \pm 0.30$ at 95\% CL.
		
		\item Our results are in excellent agreement with the official ATLAS and CMS combinations, validating both our simulation approach and the $ HPOprodMFV\_UFO$ model implementation. The consistency between our simplified simulation and the full experimental analysis demonstrates the utility of such frameworks for exploratory phenomenological studies.
		
		\item We have discussed the implications of our $\kappa_g$ constraints for various BSM scenarios, including models with new heavy particles, two-Higgs-doublet models, and composite Higgs models. The current bounds already probe new physics scales up to several hundred GeV, with much stronger constraints expected from the HL-LHC.
		
		\item The complete dataset of 26 simulated parameter points, including cross sections and statistical errors, is provided in Table~\ref{tab:full_results} and can be used for further studies or comparisons with other models.
	\end{itemize}
	
	The present work will add to the current endeavor to understand the nature of the Higgs boson and detect any indications for new physics. The methodology developed here can be extended in several directions:
	
	\begin{itemize}
		\item \textbf{Multi-dimensional fits}: By varying multiple couplings simultaneously ($\kappa_W$, $\kappa_Z$, $\kappa_b$, $\kappa_\tau$, etc.), one can explore correlated deviations that arise in specific BSM scenarios.
		
		\item \textbf{CP violation}: The $ HPOprodMFV\_UFO$ model includes parameters that control CP-violating phases. Extending our analysis to include these would allow sensitivity to CP-odd components in the $hgg$ coupling.
		
		\item \textbf{Other production modes}: While we focused on gluon fusion, the same framework can be applied to vector boson fusion (VBF), associated production (VH), and $t\bar{t}H$ production, providing complementary constraints.
		
		\item \textbf{HL-LHC projections}: Adapting our simulation to HL-LHC conditions ($\sqrt{s} = 14$ TeV, higher luminosity, improved detector performance) would provide valuable forecasts for future measurements.
		
		\item \textbf{Combination with other channels}: Including additional Higgs decay channels such as $h \to ZZ^*$ and $h \to WW^*$ would yield stronger constraints and allow cross-checks of the $\kappa_g$ dependence.
	\end{itemize}
	
	In conclusion, our study demonstrates the power of combining Monte Carlo simulations with experimental likelihood profiles to constrain Higgs couplings. The excellent agreement with ATLAS data confirms the validity of the $ HPOprodMFV\_UFO$ model and establishes a reliable framework for future phenomenological investigations. As the LHC continues to accumulate data and the HL-LHC era approaches, such studies will play an increasingly important role in guiding experimental searches and interpreting their results in terms of fundamental theory parameters.
	
	
	

\end{document}